\documentclass{epl}

\title{Orbital ordering and enhanced magnetic frustration of strained
  \chem{BiMnO_3} thin films}
\author{C.-H. Yang\inst{1} \and T. Y. Koo\inst{2} \and S.-H.
Lee\inst{1} \and C. Song\inst{1}, K.-B. Lee\inst{1} \and Y. H.
Jeong\inst{1}}

\institute{
 \inst{1}
  Department of Physics \& Electron Spin Science
Center and \ \
  \inst{2}
  Pohang Accelerator Laboratory, Pohang University of
Science and Technology, Pohang, 790-784, S. Korea}


\begin{document}

\maketitle

\begin{abstract}
Epitaxial thin films of multiferroic perovskite \chem{BiMnO_3}
were synthesized on \chem{SrTiO_3} substrates, and orbital
ordering and magnetic properties of the thin films were
investigated. The ordering of the \chem{Mn^{3+}} $e_g$ orbitals at
a wave vector $(\rm\frac{1}{4}\frac{1}{4}\frac{1}{4})$ was
detected by Mn K-edge resonant x-ray scattering. This peculiar
orbital order inherently contains magnetic frustration. While bulk
\chem{BiMnO_3} is known to exhibit simple ferromagnetism, the
frustration enhanced by in-plane compressive strains in the films
brings about cluster-glass-like properties.
\end{abstract}

Orbital degrees of freedom play a critical role in determining the
physical properties of transition-metal
oxides~\cite{Tokura,YMurakami}. \chem{BiMnO_3} with manganese
electronic configuration of $t_{2g}^{\rm 3}e_g^{\rm 1}$ is such an
example where $e_g$ orbital ordering is of central importance.
\chem{BiMnO_3} is particularly interesting, because it belongs to
so-called multiferroic materials which is currently of immense
interest~\cite{JPD}. The different degrees of freedom of the
system such as electric polarization, orbital, and magnetic spin
compete with each other, and ferroelectricity, orbital ordering,
and ferromagnetism manifest themselves~\cite{ITroyanchuk}. This
complex multiferroic behavior is already apparent from the
chemical formula; $\rm Bi^{3+}$ carries a stereochemically active
$6s^{\rm 2}$ lone pair which induces a local distortion and
bonding between Bi and neighboring oxygens~\cite{Seshadri}, while
$\rm Mn^{3+}$ is Jahn-Teller (JT) active  and has its share in
both the orbital and spin degrees of freedom. Since Mn orbital
ordering is necessarily coupled to the distortion caused by the Bi
lone pair and also determines the magnetic
interaction~\cite{Khomskii}, the detailed knowledge of the orbital
ordering would be essential to the understanding of the
multiferroic system \chem{BiMnO_3}.

Despite recent results regarding \chem{BiMnO_3} such as first
principle calculations \cite{Hill,TShishidou}, neutron diffraction
\cite{AMoreira,TAtou}, magnetocapacitance effects \cite{TKimura},
and thin film properties \cite{AMoreiraFilm,ASharan,Eerenstein},
the orbital ordering pattern and its influences on the
ferromagnetic and/or ferroelectric properties await detailed
experimental investigations. One factor hindering progress has
been the lack of quality samples in bulk or thin film form.
However, we have succeeded in obtaining high quality, epitaxial
{\it thin films} and this allowed us to address the issues related
to the orbital ordering. In this article, we wish to accomplish
two major objectives: first, we will present the resonant x-ray
scattering (RXS) data for epitaxial films  as a {\it direct}
evidence which confirms the rather peculiar orbital ordering
pattern for \chem{BiMnO_3} derived previously  by Atou et al. from
neutron diffraction refinements~\cite{TAtou}. Second, we wish to
demonstrate the magnetic consequences of the orbital order, that
is, the {\em frustration} inherent in \chem{BiMnO_3} due to its
orbital order manifest itself in the magnetic properties of
strained thin films.

\begin{figure}
\vspace{-5.3cm} \onefigure[width=13cm]{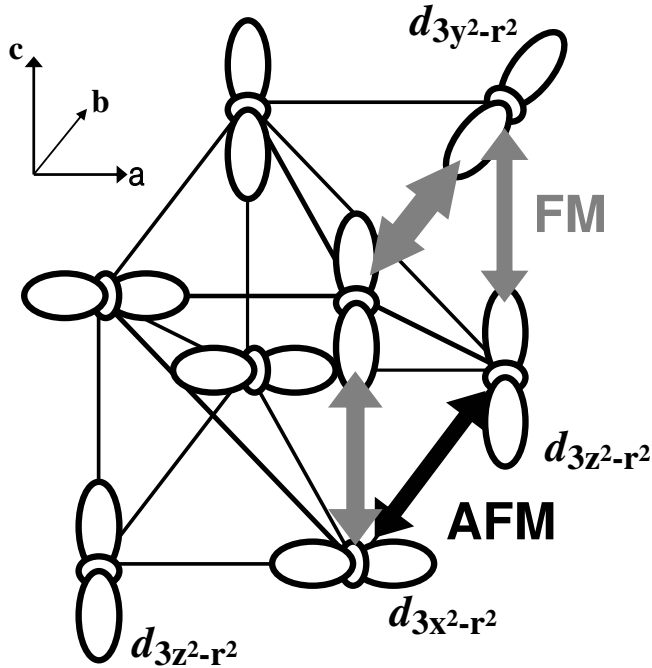} \vspace{-3cm}
\caption{The Mn $e_g$ orbitals in \chem{BiMnO_3} are shown; Bi and
O ions are omitted. The actual structure of \chem{BiMnO_3} is
monoclinic with space group $C2$. The superexchange interactions
indicated by gray and black arrows are ferromagnetic and
antiferromagnetic, respectively. Notice that frustration occurs in
the plaquette. The thickness of the arrows implies the strength
estimated from ref.~\cite{ExchangeInteraction}.} \label{BMOORBIT}
\end{figure}

From the refinement of the neutron powder diffraction data at room
temperature  the space group (monoclinic $C2$) and local bond
lengths and angles of \chem{BiMnO_3} were all determined
previously~\cite{AMoreira,TAtou}; these structural data contain
information on  the orientation of JT-distorted \chem{MnO_6}'s,
from which the compatible ordered pattern of the $e_g$ orbitals of
\chem{Mn^{3+}} ions was inferred by Atou et al.. The orbital order
of \chem{BiMnO_3}, illustrated schematically in
Fig.~\ref{BMOORBIT}, is rather peculiar in the sense that each
plane parallel to (111) contains only one kind of the $e_g$
orbital within the plane and the stacking sequence goes as
$d_{3z^2-r^2}$/$d_{3x^2-r^2}$/$d_{3z^2-r^2}$/$d_{3y^2-r^2}$/$d_{3z^2-r^2}$/$d_{3x^2-r^2}$/$\ldots$.
Thus a superlattice modulation would occur at wavevector
$(\rm\frac{1}{4}\frac{1}{4}\frac{1}{4}$) (in cubic reciprocal
lattice unit; the subscript $m$ is attached to monoclinic
indices.). It is noted that the expected orbital order of
\chem{BiMnO_3}  is rather different from that of a similar
compound ${\rm LaMnO_3}$ at ($\rm\frac{1}{2}\frac{1}{2}0$)
\cite{Song}.

According to Goodenough, the superexchange interaction between
\chem{Mn^{3+}} ions becomes antiferromagnetic (AFM) when both
$e_g$ orbitals point perpendicular to the bond direction, while
the ferromagnetic (FM) interaction prevails when  one of the $e_g$
orbitals points along the bond direction~\cite{Goodenough}. The
striking consequence of the rule in this case is the presence of
frustration as indicated in Fig.~1.  In bulk \chem{BiMnO_3},
however, the FM interactions seem to dominate over the AFM ones
and consequently the normal FM ground state
results~\cite{AMoreira}. Since thin films are generally strained
and the superexchange interaction depends sensitively on the
inter-ion distance,
%~\cite{DBloch},
a possibility exists that
strains drive the system to a point where the strength of the FM
and AFM interactions are balanced ~\cite{ExchangeInteraction}.
This then suggests that frustration come into operation and induce
a frustrated state in \chem{BiMnO_3} thin films. Thus magnetic
properties of thin films are of keen interest.

\chem{BiMnO_3} films were grown on ${\rm SrTiO_3}$ substrates from
a bulk target via pulsed laser deposition. The substrate
temperature was maintained at 460${\rm ^o}$C during deposition and
the oxygen partial pressure was kept at 4~${\rm
mTorr}$~\cite{Eerenstein}. Note that the growth temperature of
460${\rm ^o}$C is rather low compared to 700$\sim$750${\rm ^o}$C
adopted by some other groups~\cite{AMoreiraFilm,ASharan}. While
these groups used targets with excessive Bi contents to compensate
for Bi evaporation during film growth, the low deposition
temperature we used presented an advantage of suppressing Bi
evaporation and led to a single phase, epitaxial films. X-ray
absorption measurements proved that the Mn ions attain the valence
of 3+ and SEM-EDX indicated that Bi:Mn = 1:1 within its
resolution. The surface roughness measured by atomic force
microscopy was $\sim$5 \AA. The thickness of the films was
estimated from X-ray reflectivity measurements.

\begin{figure}[h]
\vspace{-5cm} \hspace{-1cm} \onefigure[width=14.4cm]{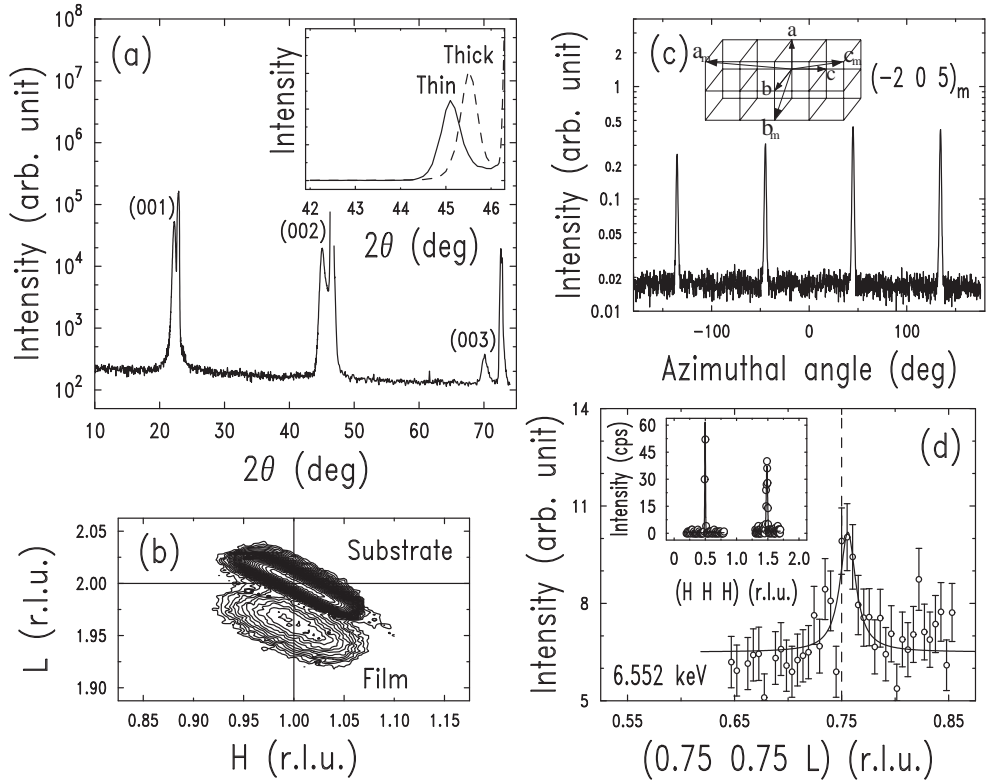}
\vspace{-1cm} \caption{X-ray diffraction results for
\chem{BiMnO_3} films on $\rm SrTiO_3$(001) substrates.
  (a) $\theta-2\theta$ scan of a film with thickness 40 nm.
  Inset illustrates the (002) peak for two films of
  different thickness (solid line, 40 nm; dashed line, 70 nm).
  (b) Reciprocal space mapping of the 70 nm film around the ${\rm SrTiO_3}$ (102) peak.
 The tetragonality factor $t\,=\,{\rm +1.5~\%}$ with
  $c-$axis elongation and in-plane compression.
  (c) Monoclinic $(\bar{2}05)_{m}$ peak which is $31.2^{\rm o}$
  off from the surface normal shows a four-fold symmetry from twinning under azimuthal rotation.
  Inset describes the relationship between the monoclinic and simple cubic cell.
  (d) Resonant x-ray scattering peak at Mn K-absorption edge from the 70 nm film.
 The line is guide to the eye.
  Inset displays non-resonant superlattice reflections at
  $\rm (\frac{1}{2}\frac{1}{2}\frac{1}{2})$ and
  $\rm (\frac{3}{2}\frac{3}{2}\frac{3}{2})$.} \label{BMOXRAY}
\end{figure}

Fig. 2 summarizes the x-ray diffraction results at room
temperature for two films with thickness approximately 40 nm and
70 nm deposited on \chem{SrTiO_3}(001) substrates.  X-ray
diffraction measurements were carried out using synchrotron
radiation of the Pohang Light Source (PLS). The $\theta$-$2\theta$
scan of the film with thickness 40 \un{nm} is presented in
Fig.~\ref {BMOXRAY}(a). The inset illustrates that the 40 nm film
has a larger out-of-plane ($c$-axis) lattice parameter than that
of the film with thickness 70 nm. Fig.~\ref{BMOXRAY}(b) is the
mapping of the (102) peak of the 70 nm film, and shows that the
film has an epitaxial relationship with the substrate. The
in-plane lattice constant $a$ is obtained from the Bragg peak
position. The films are under in-plane compression and
out-of-plane expansion; the tetragonal strain factor, defined as
$t\equiv (c-a)/a$, is +1.5~$\%$ for the thick film (70 nm) and
+2.8~$\%$ for the thin one (40 nm). The in-plane compression would
strengthen the AFM interactions depicted in Fig.~1, and thus
enhance frustration.

While tetragonality was adopted to describe the \chem{BiMnO_3}
films, their exact symmetry is still monoclinic (space group $C2$)
with the unique two-fold axis (monoclinic $b$-axis) lying in the
surface plane. The determination of the symmetry was carried out
by searching for several specific monoclinic reflections from the
films. A representative Bragg peak $\rm (\overline{2}05)_m$ of the
70 nm film is displayed in Fig.~\ref{BMOXRAY}(c); the four peaks
appearing as a function of azimuthal angle indicate the existence
of naturally occurring 4-fold twins. In the inset of
Fig.~\ref{BMOXRAY}(c) illustrated is the monoclinic unit cell of
\chem{BiMnO_3}, which is related to the cubic one via the
transformation matrix, $\rm (a~b~c)_m =
(a~b~c)$(1~-1~1/1~1~1/-2~0~2). The monoclinic lattice parameters
were determined as: $a_m\rm=9.820~\AA$, $b_m\rm=5.441~\AA$,
$c_m\rm=9.852~\AA$, $\beta\rm=110.0^o$ for the 40 nm film, and
$a_m\rm=9.774~\AA$, $b_m\rm=5.440~\AA$, $c_m\rm=9.790~\AA$,
$\beta\rm=109.1^o$ for the 70 nm one. The monoclinic structure
also produces superlattice reflections at
($\rm\frac{1}{2}\frac{1}{2}\frac{1}{2}$) and
($\rm\frac{3}{2}\frac{3}{2}\frac{3}{2}$) (corresponding to
monoclinic (002)$_m$ and (006)$_m$, respectively)  as shown in the
inset of Fig.~\ref{BMOXRAY}(d).

\begin{figure}
\onefigure[height=9cm,angle=90]{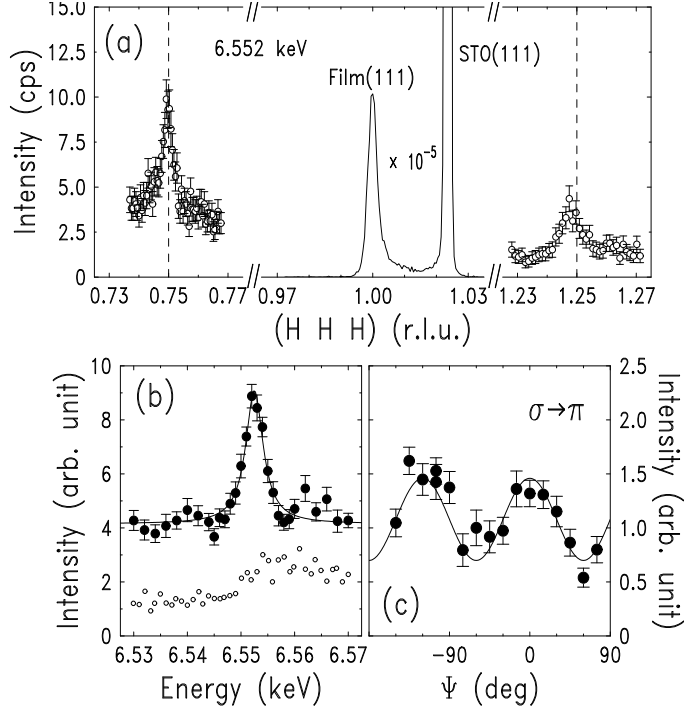} \caption{Resonant X-ray
scattering results for the $\rm BiMnO_3$ film of thickness 40 nm
on a \chem{SrTiO_3} (111) substrate.
  (a) Reciprocal lattice scans along the (111)-direction. The
  (111) peaks %from the film and substrate
  are normal ones, while the peaks at
  $\rm (\frac{3}{4}\frac{3}{4}\frac{3}{4})$ and
  $\rm (\frac{5}{4}\frac{5}{4}\frac{5}{4})$ are resonant ones.
  Notice the intensity difference.
  (b) The integrated intensity of the $\rm (\frac{3}{4}\frac{3}{4}\frac{3}{4})$ peak as a
   function of photon energy. The line is a fit to the Lorentzian
   shape. Also plotted is the fluorescence data (open circles).
 (c) Azimuthal angle ($\Psi$) dependence of the integrated intensity
  of the $\rm (\frac{3}{4}\frac{3}{4}\frac{3}{4})$
  peak. The intensity is due to scattering from $\sigma$- to $\pi$-polarization. $\Psi = 0$ is set when the scattering plane
  coincides with the $\rm (1\bar{1}0)$ plane. The line is a sine function with three-fold symmetry.} \label{BMOSL}
\end{figure}

With the monoclinicity of the films established, let us focus our
attention on superlattice peaks from orbital ordering. In this
regard, it is worth noting that normal charge scattering due to
the ($\rm\frac{1}{4}\frac{1}{4}\frac{1}{4}$) modulation (or
equivalently (001)$_m$) is not completely forbidden, but strongly
suppressed in the monoclinic structure. However, the suppression
was large enough to allow RXS due to orbital ordering to be
observed. Procedures for RXS measurements at PLS were described
previously~\cite{Song}; the measurements were carried out at the
Mn K-absorption edge of photon energy 6.552 keV at room
temperature. Fig.~\ref{BMOXRAY}(d) depicts RXS results for the 70
nm film, and a peak due to the
($\rm\frac{1}{4}\frac{1}{4}\frac{1}{4}$) modulation is clearly
seen. Although the peak occurs at the expected position, this fact
alone is not sufficient to conclude that it is indeed due to
orbital ordering; in addition, azimuthal variation of the RXS
intensity with respect to the scattering vector and the
polarization dependence need to be checked.

Since the [111]-direction of the film grown on a ${\rm
SrTiO_3}$(001) substrate was not parallel to the surface normal,
azimuthal scans turned out to be difficult to perform. Thus for
further RXS measurements a new sample of approximate thickness 40
nm was grown on ${\rm SrTiO_3}$(111). The RXS results for the new
sample are shown in Fig.~\ref{BMOSL}. Fig.~\ref{BMOSL}(a) shows
that the expected superlattice peaks are located at the correct
positions as satellites of the (111) main peak. The main peak in
the figure was obtained from normal scattering, while the side
ones are from resonant scattering. Fig.~\ref{BMOSL}(b) illustrates
the resonant nature of the latter peaks. The integrated intensity
of the ($\rm\frac{3}{4}\frac{3}{4}\frac{3}{4}$) peak rises sharply
to a maximum at 6.552 keV with a width of about 8 eV, and an
increase of fluorescence indicates the  Mn K-absorption edge. Note
that normal charge scattering persists at off-resonance energies.
Since only RXS allows $\sigma$ to $\pi$ scattering (here $\sigma$
and $\pi$ indicate the polarization perpendicular and parallel to
the scattering plane, respectively), we carried out azimuthal
scans with the $\sigma$ to $\pi$ channel allowed with a polarizer
at the detector side. Fig.~\ref{BMOSL}(c) displays the azimuthal
angle dependence of the $\sigma$ to $\pi$ intensity of the
($\rm\frac{3}{4}\frac{3}{4}\frac{3}{4}$) peak normalized to that
of the (111) fundamental peak; a conspicuous intensity oscillation
is seen. The azimuthal angle dependence is the hallmark of RXS
from orbital ordering and the thorough analysis including twin
effects, the details of which are presented elsewhere~\cite{RXS},
yielded the solid line in the figure.

\begin{figure}
\vspace{-1cm} \onefigure[width=12cm]{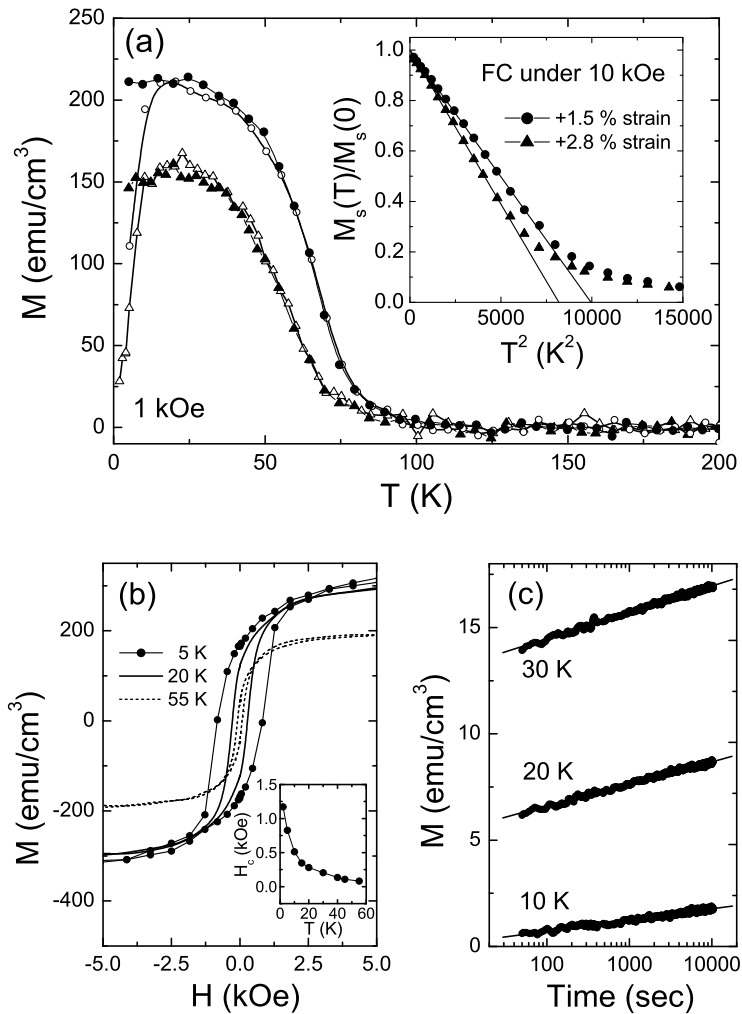} \vspace{-1cm}
\caption{Magnetic properties of \chem{BiMnO_3} films on
\chem{SrTiO_3}(001). (a)  ZFC (open symbol) and FC (solid symbol)
magnetization of the strained films ($t\,=\,$+1.5~\% and +2.8~\%)
is shown as a function of temperature. The in-plane field strength
was 1 kOe. Inset displays that the saturation magnetization
decreases with the $T^2$ behavior. (b) $M-H$ hysteresis curves of
the film with $t\,=\,\rm +1.5~\%$ at selected temperatures. Inset
shows the $T$ dependence of the coercive field $H_c$. (c) $M$ of
the film with $t\,=\,\rm +2.8~\%$ is plotted as a function of
time. It was zero-field cooled and then a field of 100 Oe was
applied. Notice the slow dynamics (time in log scale) with slope
increasing with $T$.} \label{BMOMAG}
\end{figure}

We now turn to the second major point, that is, the magnetic
properties of the \chem{BiMnO_3} films; magnetic properties were
measured employing a Quantum Design MPMS. In Fig.~\ref{BMOMAG}(a),
the magnetization $M$ for the two strained films on
\chem{SrTiO_3}(001) substrates ($t\,=\,$+1.5, +2.8 \% for the 70
and 40 \un{nm} thick films, respectively) is plotted as a function
of temperature $T$. When $M$ is measured with an in-plane field of
1 kOe (FC, field-cooled), $M$ increases below a certain
temperature, which depends on strain (or thickness) of the film
and is much lower than the bulk transition temperature 105 K, as
$T$ is reduced. In addition, the thin films possess smaller
saturation moments, when measured under 4 T at 5 K, of 2.47~$\mu_B
/{\rm Mn}$ ($t\,=\,$+1.5 \% film) and 2.12~$\mu_B /{\rm Mn}$
($t\,=\,$+2.8 \% film) compared to the bulk value of 3.6 $\mu_B
/{\rm Mn}$ \cite{AMoreira}. More striking is the fact that $M$
measured with the same field value but first cooled under zero
field (ZFC, zero field-cooled) shows deviation from the FC values
at low temperatures. The inset of Fig.~\ref{BMOMAG}(a) shows that
the saturated magnetization of both 40 and 70 \un{nm} films,
measured at 10 kOe, decreases with temperature displaying the
$T^2$ behavior instead of the well-known $T^{3/2}$ law
\cite{UKobler}. Fig.~\ref{BMOMAG}(b) is the plot of $M-H$ curves
for the 70 nm film at selected temperatures. In obtaining these
data, magnetic field was applied parallel to the [100] or [010]
direction. These measurements were performed after zero-field
cooling from 150 K, at which the film was paramagnetic and didn't
show any thermal and field hysteretic behavior. Notice that the
coercive field ($H_c$) rapidly increases below the blocking (or
freezing) temperature ($T_b$ $\sim$ 15 K). $T_b$ also indicates
the splitting point of ZFC and FC magnetizations in
Fig.~\ref{BMOMAG}(a).

The magnetic features of the \chem{BiMnO_3} films displayed in
Fig.~\ref{BMOMAG}(a) and (b), a clear difference in FC and ZFC $M$
values and a rapid increase of $H_c$ below $T_b$, are undoubtedly
not those of a typical ferromagnet but of a random magnetic
system. Thus to further clarify the magnetic state we sought for
time dependent behaviors in $M$ usually appearing in random
systems. After ZFC a magnetic field of 100 Oe was applied and $M$
was measured as a function of time. Fig.~\ref{BMOMAG}(c)
illustrates that $M$ of the \chem{BiMnO_3} film shows a
logarithmic behavior in time, and that this relaxational dynamics
gets faster with increasing $T$. Similar features are often seen
in disordered oxide materials and termed
'cluster-glass-like'~\cite{disorder1,disorder2}. It is rather
remarkable that these unusual magnetic properties are found in
\chem{BiMnO_3} films without disorder. In this regard it may be
worth noting that epitaxial magnetite thin films were also found
to possess frustration leading to a magnetic state different from
the bulk one~\cite{magnetite}.

While the detailed account of the magnetic properties of
\chem{BiMnO_3} films certainly requires further efforts, it may
still be concluded that the inherent frustration of \chem{BiMnO_3}
is enhanced in films with positive tetragonal strain ($t>0$) and
this enhanced frustration alters the FM ground state seen in
bulks.  Borrowing the theory of superparamagnetism, one may
estimate the magnetic cluster size from the measured $T_b$ and
$H_c$ at zero temperature~\cite{superpara}. The characteristic
size estimated in this way is about 3 or 4 \un{nm}, and this is
sufficiently smaller than the correlation length of the orbital
order $\sim$30 \un{nm}. The latter length is extracted from the
width of the x-ray superlattice peak using the Scherrer formula.
The clear separation of the two length scales indicates that the
anomalous magnetic behavior of the \chem{BiMnO_3} films is an
intrinsic phenomenon, rather than extrinsic one induced by, for
example, finite size domains. In conclusion we thank the financial
supports from the SRC program of MOST/KOSEF.

\end{document}